\batchmode
\makeatletter
\def\input@path{{\string"C:/Users/Steven/Google Drive/NCR/Publications/[TRO2018] A Switched Systems Approach to Tracking Control with Intermittent Feedback/arXiv/\string"}}
\makeatother
\documentclass[english,conference]{IEEEtran}
\usepackage[T1]{fontenc}
\usepackage[latin9]{inputenc}
\usepackage{array}
\usepackage{multirow}
\usepackage{amsmath}
\usepackage{amsthm}
\usepackage{amssymb}
\usepackage{graphicx}

\makeatletter

\providecommand{\tabularnewline}{\\}

\theoremstyle{definition}
\newtheorem{assumption}{Assumption}
  \theoremstyle{plain}
  \newtheorem{thm}{\protect\theoremname}
  \theoremstyle{remark}
  \newtheorem{rem}{\protect\remarkname}

\usepackage{fouridx}
\usepackage{cite}
\IEEEoverridecommandlockouts

\makeatother

\usepackage{babel}
\providecommand{\remarkname}{Remark}
\providecommand{\theoremname}{Theorem}

\begin{document}

\title{A Switched Systems Approach to Path Following with Intermittent State
Feedback\thanks{Hsi-Yuan Chen, Zachary I. Bell, Patryk Deptula, and Warren E. Dixon
are with the Department of Mechanical and Aerospace Engineering,University
of Florida, Gainesville, Florida, 32611-6250, USA. Email: \{hychen,bellz121,pdeptula,wdixon\}@ufl.edu.}}

\author{Hsi-Yuan Chen, Zachary I. Bell, Patryk Deptula, Warren E. Dixon}
\maketitle
\begin{abstract}
Autonomous agents are often tasked with operating in an area where
feedback is unavailable. Inspired by such applications, this paper
develops a novel switched systems-based control method for uncertain
nonlinear systems with temporary loss of state feedback. To compensate
for intermittent feedback, an observer is used while state feedback
is available to reduce the estimation error, and a predictor is utilized
to propagate the estimates while state feedback is unavailable. Based
on the resulting subsystems, maximum and minimum dwell time conditions
are developed via a Lyapunov-based switched systems analysis to relax
the constraint of maintaining constant feedback. Using the dwell time
conditions, a switching trajectory is developed to enter and exit
the feedback denied region in a manner that ensures the overall switched
system remains stable. A scheme for designing a switching trajectory
with a smooth transition function is provided. Simulation and experimental
results are presented to demonstrate the performance of control design.
\end{abstract}


\section{Introduction\label{sec:Introduction}}

Acquiring state feedback is at the core of ensuring stability in control
designs. However, factors such as the task definition, operating environment,
or sensor modality can result in temporary loss of feedback. For example,
agents may be required to limit communication during predefined time
frames or when traversing through certain regions. Motivated by such
factors, various path planning and control methods have been developed
that seek to ensure uninterrupted feedback (cf., \cite{gans2003ptv,Hutchinson96,Gans2011,GansT.A.a,Hu.Gans.ea2010,Hu2010a,Hu2009a,Chen2007,Chen.Dawson.ea2005,Palmieri2012,GansSwitchVS2007,ChesiSwVS04}).
Such results inherently constrain the trajectory or behavior of a
system. For instance, visual servoing applications for nonholonomic
systems can result in limited, sharp-angled or non-smooth trajectories
to keep a target in the camera field-of-view (FOV) as illustrated
in results such as \cite{Gans.Hutchinson2007,Mariottini2007,Lopez-Nicolas.Gans.ea2010}.
Rather than trying to constrain the system to ensure continuous feedback
is available, the approach in this paper leverages switched systems
methods to achieve an objective despite intermittent feedback. 

Solutions to relaxing the constant feedback constraint have been investigated.
For example, methods to relax the requirement of keeping landmarks
in the FOV have been developed in results such as \cite{Mehta2008a}
and \cite{Jia2015}. In \cite{Mehta2008a}, multiple landmarks are
linked together by a daisy-chaining approach where new landmarks are
mapped onto the initial world frame and are used to provide state
feedback when initial landmarks leave the FOV. Similar concepts were
adopted in \cite{Jia2015}, where a wheeled mobile robot (WMR) is
allowed to navigate around a landmark without constantly keeping it
in the FOV by relating feature points in the background to the landmark
and thus provide state feedback. Although the objective to eliminate
the requirement of constant visual on the landmark is achieved, state
feedback is assumed to be available during periods when the landmark
is outside the FOV. Such daisy-chaining approaches provide state feedback
in an ideal scenario, but the accuracy of the feedback may degrade
or even diverge in the presence of measurement noise and disturbances
in the dynamics.

Conventional approaches to the simultaneous localization and mapping
(SLAM) problem, such as the works in \cite{Klein.Murray2007,Davison.Reid.ea2007,Cremers2017},
use relationships between features or landmarks to estimate the pose
(i.e., position and orientation) of the sensor, usually a monocular
camera, and simultaneously determine the position of landmarks with
respect to the world frame. Typically, a feature rich environment
with sufficient measurements are required for SLAM methods to provide
state estimation. However, a well-known drawback with SLAM algorithms
is that without proper loop closures the estimates will drift over
time due to the accumulation of measurement noise (cf., \cite{Williams.Cummins.ea2009,Cadena.Carlone.ea2016}).
In this paper, a state estimate dynamic model propagates the state
estimate when feedback is not available, and no additional feedback
information is required. Sufficient conditions may be derived via
a Lyapunov-based analysis to ensure the loop closures are achieved
before the state estimates degrade beyond a desired threshold. 

Stability of systems that experience random state feedback has been
analyzed in previous literature. Typically, the intermittent loss
of measurement is modeled as a random Bernoulli process with a known
probability. Resulting trajectories are then analyzed in a probabilistic
sense, where the expected value of the estimation error is shown to
converge asymptotically, compared to the result in this paper which
examines the behavior of the actual tracking and estimation errors. 

The networked control systems (NCS) community has also examined systems
with temporarily unavailable measurements. Results such as \cite{Garcia2011,Mehta2013,Garcia2013,McCourt.Garcia.ea2014}
rely on a decision maker that is independent of the estimator or controller
to determine when to broadcast sensor information. The objective in
these results is to minimize the cost of network bandwidth by reducing
the frequency of data transmission. In \cite{Leonard2013,Liang2011,Shi2009},
data loss is modeled as random missing outputs and noisy measurements.
In each case, state estimates are propagated by a model of the controlled
system during the periods when transmission is missing. On the contrary,
the availability of sensor information in this paper is not controlled
by a decision maker but instead determined by the region in which
the actual states are located. Therefore, sensor information is only
available when the states are inside a feedback region.

It is well known that slow switching between stable subsystems may
result in instability as explained in \cite{Liberzon2003}. For slow
switching between stable subsystems, the underlying strategy for proving
stability involves developing switching conditions to ensure the overall
system is stable. If a common Lyapunov function exists for all subsystems
such that the time derivative of the Lyapunov function is upper bounded
by a common negative definite function, the overall system is proven
to be stable in \cite{Liberzon2003}. For cases where a common Lyapunov
function cannot be determined, multiple subsystem-specific Lyapunov
functions are used. In general, the overall Lyapunov function is discontinuous
and jumps may occur at switching interfaces. Therefore, the stability
of such a system is achieved by placing switching conditions on the
subsystems to enforce a decrease in the subsystem-specific Lyapunov
functions between each successive activation of the respective subsystems.
Typically, these requirements manifest as (average) dwell time conditions
which specifies the duration for which each subsystem must remain
active, as described in \cite{Liberzon2003}. 

When a subset of the subsystems is unstable, a layer of complication
is introduced to the analysis. A stability analysis is provided in
\cite{Zhai2001} for switched systems with stable and unstable linear
time invariant (LTI) subsystems, where an average dwell time condition
is developed. Similarly, the authors in \cite{Mueller.Liberzon2012}
developed dwell time conditions for nonlinear switched systems with
exponentially stable and unstable subsystems. However, dwell time
conditions typically require the stable subsystems to be activated
longer than the unstable subsystems, as indicated in \cite{Zhai2001}.
In \cite{Parikh.Cheng.ea2017}, the authors developed an observer
to estimate the depths of feature points in a image from a monocular
camera and use a predictor to propagate the state estimates when the
features are occluded or outside the FOV. Based on the error system
formulation, the subsystem for the observer is stable, while the subsystem
for the predictor is unstable. An average dwell time condition is
developed to ensure the stability of the switched system. However,
the focus of \cite{Parikh.Cheng.ea2017} is the estimation of feature
depths and therefore have not focused on achieving a control objective
when feedback is unavailable. 

The development in this paper aims to achieve a path following objective
despite intermittent loss of feedback. The novelty of this result
is guaranteeing the stability of following a path which lies outside
a region with feedback while maximizing the amount of time the agent
spends in the feedback-denied environment. Switched systems methods
are used to develop a state estimator and predictor when state feedback
is available or not, respectively. Since switching occurs between
a stable subsystem when feedback is available and an unstable subsystem
when feedback is not available, dwell time conditions are developed
that determine the minimum time that the agent must be in the feedback
region versus the maximum time the agent can be in the feedback denied
region. Using these dwell time conditions, a switching trajectory
is designed based on the dwell time conditions that leads the agent
in and out of the feedback denied region so that the overall system
remains stable. The most similar result to this paper is in \cite{Chen.Bell.ea2017},
which includes state prediction and control for a nonholonomic system
moving around an obstacle. The goal in \cite{Chen.Bell.ea2017} is
to regulate a nonholonomic vehicle to a set-point in the presence
of intermittent feedback. However, the difficulty of path following
in the current paper arises when the system is outside the feedback
region. 

The paper is organized as follows. In Section \ref{sec:Kinematic-Motion-Model},
the system model is introduced. In Section \ref{sec:Objective}, the
tracking and estimation objective is given and the respective error
systems are defined. Based on the error dynamics, a Lyapunov-based
stability analysis for the resulting switched system is performed
in Section \ref{sec:Switched-System-Analysis} to develop the dwell
time conditions and to show stability of the overall system. In Section
\ref{sec:Trajectory-Design}, a strategy for designing a switching
trajectory is presented. A simulation is provided in Section \ref{sec:Simulation}
and an experiment is provided in Section \ref{sec:Experiments} to
demonstrate the performance of the approach. 

\section{System Model\label{sec:Kinematic-Motion-Model}}

Consider a dynamic system subject to an exogenous disturbance as

\begin{equation}
\dot{x}(t)=f(x(t),t)+v(t)+d(t),\label{eq:qDot dynamics}
\end{equation}
where $x(t),\:\dot{x}(t)\in\mathbb{R}^{n}$ denote a generalized state
and its time derivative, $f:\:\mathbb{R}^{n}\times\mathbb{R}\rightarrow\mathbb{R}^{n}$
denotes the locally Lipschitz drift dynamics, $v(t)\in\mathbb{R}^{n}$
is the control input, and $d(t)\in\mathbb{R}^{n}$ is the exogenous
disturbance where the Euclidean norm is bounded as $\Vert d(t)\Vert\leq\bar{d}\in\mathbb{R}_{\geq0}$
with $n\in\mathbb{N}$ and $t\in\mathbb{R}_{\geq0}$. 

\section{State Estimate and Control Objective\label{sec:Objective}}

In this paper, the overall objective is to achieve path following
under intermittent loss of feedback. Specifically, a known feedback
region is denoted as a closed set $\mathcal{F}\subset\mathbb{R}^{n}$,
where the complement region where feedback is unavailable is denoted
by $\mathcal{F}^{c}$. That is, feedback is available when $x(t)\in\mathcal{F}$
and unavailable when $x(t)\in\mathcal{F}^{c}$. 

A desired path is denoted as $x_{d}\subset\mathcal{F}^{c}$. It is
clear that state feedback is unavailable while attempting to follow
$x_{d}$, and hence the system must return to the feedback region
$\mathcal{F}$ intermittently to maintain stability. Therefore, a
switching trajectory, denoted by $\bar{x}_{d}(t)\in\mathbb{R}^{n}$,
is designed to overlay $x_{d}$ as much as possible while adhering
to the subsequently developed dwell time constraints. To quantify
the ability of the controller to track the switching trajectory, the
tracking error $e(t)\in\mathbb{R}^{n}$ is defined as

\begin{eqnarray}
e(t) & \triangleq & e_{1}(t)+e_{2}(t),\label{eq:e}
\end{eqnarray}
where the estimate tracking error $e_{1}(t)\in\mathbb{R}^{n}$ is
defined as

\begin{eqnarray}
e_{1}(t) & \triangleq & \hat{x}(t)-\bar{x}_{d}(t),\label{eq:e1}
\end{eqnarray}
and the state estimation error $e_{2}(t)\in\mathbb{R}^{n}$ is defined
as

\begin{eqnarray}
e_{2}(t) & \triangleq & x(t)-\hat{x}(t),\label{eq:e2}
\end{eqnarray}
where $\hat{x}(t)\in\mathbb{R}^{n}$ is the state estimate.

Based on (\ref{eq:e1}) and (\ref{eq:e2}), the control objective
is to ensure that $e_{1}(t)$ and $e_{2}(t)$ converge, and therefore
$e(t)$ will converge. To facilitate the subsequent development, let
the composite error vector be defined as $z(t)\triangleq\left[\begin{array}{cc}
e_{1}^{T}(t) & e_{2}^{T}(t)\end{array}\right]^{T}$.

\section{Controller and Update Law Design\label{sec:Controller}}

To facilitate the subsequent analysis, two subsystems are defined
to indicate when the states are inside or outside the feedback region.
When $x(t)\in\mathcal{F}$, an exponentially stable observer can be
designed using various approaches (e.g., observers such as \cite{Parikh.Cheng.ea2017,DaniT.A.b,Bell.Chen.ea2017}
could be used). The subsequent development is based on an observer
update law designed as\footnote{Once $x(t)\in\mathcal{F}$, a simple reset scheme (i.e. setting $\hat{x}(t)=x(t)$)
could be used. The reset scheme would eliminate the subsequently developed
minimum dwell time for which $x(t)$ is required to remain in the
feedback region $\mathcal{F}$. However, the subsequent development
is based on the continued use of the observer to illustrate a more
general stability condition for systems that require an observer or
$\dot{\hat{x}}(t)\in\mathcal{L}_{\infty},\ \forall t$ .}

\begin{equation}
\dot{\hat{x}}(t)=f(\hat{x}(t),t)+v(t)+v_{r}(t),\label{eq: observer dynamics}
\end{equation}
where $v_{r}(t)\in\mathbb{R}^{n}$ is a high-frequency sliding-mode
term designed as \footnote{{\small{}In cases where a piecewise-continuous controller is required,
the robustifying term in (\ref{eq: observer dynamics}) may be designed
as $v_{r}(t)=k_{2}e_{2}+\frac{\bar{d}^{2}}{\epsilon}e_{2},$ where
$\epsilon\in\mathbb{R}_{>0}$ is a design parameter.}}

\begin{equation}
v_{r}(t)=k_{2}e_{2}(t)+\bar{d}sgn(e_{2}(t)),\label{eq:vr}
\end{equation}
 where $k_{2}\in\mathbb{R}^{n\times n}$ is a constant, positive definite
gain matrix. When $x(t)\in\mathcal{F}^{c}$, the state estimate is
updated by a predictor designed as 

\begin{equation}
\dot{\hat{x}}(t)=f(\hat{x}(t),t)+v(t).\label{eq:predictor dynamics}
\end{equation}

Since the state is required to transition between $\mathcal{F}$ and
$\mathcal{F}^{c}$, a switched systems analysis is used to investigate
the stability of the overall switched system. To facilitate this analysis,
the error systems for $e_{1}(t)$ and $e_{2}(t)$ are expressed as 

\begin{align}
\dot{e}_{1}(t) & =f_{1p}\left(\bar{x}_{d}(t),\hat{x}(t),t\right),\label{eq:e1Dot_open}\\
\dot{e}_{2}(t) & =f_{2p}\left(x(t),\hat{x}(t),t\right),\label{eq:e2Dot_open}
\end{align}
where $f_{1p},\ f_{2p}:\mathbb{R}^{n}\times\mathbb{R}^{n}\times\mathbb{R}_{\geq0}\rightarrow\mathbb{R}^{n},\;p\in\{a,u\}$,
$a$ is an index for subsystems with available feedback, and $u$
is an index for subsystems when feedback is unavailable. Based on
(\ref{eq:e1Dot_open}) and the subsequent stability analysis, the
controller is designed as

\begin{equation}
v(t)=\begin{cases}
\dot{\bar{x}}_{d}(t)-f(\hat{x}(t),t)-k_{1}e_{1}(t)-v_{r}(t), & p=a,\\
\dot{\bar{x}}_{d}(t)-f(\hat{x}(t),t)-k_{1}e_{1}(t), & p=u,
\end{cases}\label{eq:u}
\end{equation}
where $\dot{\bar{x}}_{d}(t)\in\mathbb{R}^{n}$, and $k_{1}\in\mathbb{R}^{n\times n}$
is a constant, positive definite gain matrix. By taking the time derivative
of (\ref{eq:e1}) and substituting (\ref{eq: observer dynamics}),
(\ref{eq:predictor dynamics}) and (\ref{eq:u}) into the resulting
expression, (\ref{eq:e1Dot_open}) can be expressed as

\begin{equation}
\dot{e}_{1}(t)=-k_{1}e_{1}(t),\ \forall p.\label{eq:e1Dot_closed}
\end{equation}
After taking the time derivative of (\ref{eq:e2}) and substituting
(\ref{eq:qDot dynamics}), (\ref{eq: observer dynamics}) and (\ref{eq:predictor dynamics})
into the resulting expression, the family of systems in (\ref{eq:e2Dot_open})
can be expressed as

\begin{equation}
\dot{e}_{2}(t)=\begin{cases}
f(x(t),t)-f(\hat{x}(t),t)+d(t)\\
-\bar{d}sgn(e_{2}(t))-k_{2}e_{2}(t), & p=a,\\
f(x(t),t)-f(\hat{x}(t),t)+d(t), & p=u.
\end{cases}\label{eq:e2Dot closed loop}
\end{equation}

\section{Switched System Analysis\label{sec:Switched-System-Analysis}}

\begin{figure}
\includegraphics[width=1\columnwidth]{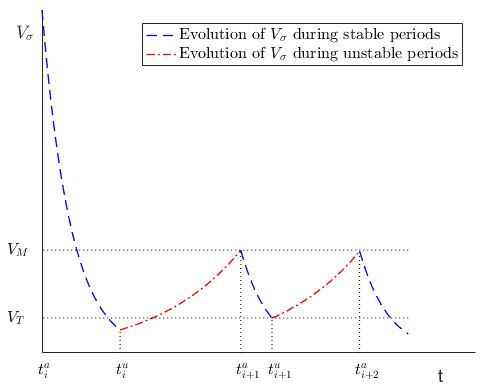}\caption{Representative illustration for the evolution of $V_{\sigma}$ during
the interval $\left[t_{i}^{a},t_{i+2}^{a}\right]$.\label{fig:evolution_of_v}}
\end{figure}

To further facilitate the analysis for the switched system, let $t_{i}^{a}\in\mathbb{R}_{\geq0}$
denote the time of the $i^{th}$ instance when $x(t)$ transitions
from $\mathcal{F}^{c}$ to $\mathcal{F}$, and $t_{i}^{u}\in\mathbb{R}_{>0}$
denote the time of the $i^{\text{th}}$ instance when $x(t)$ transitions
from $\mathcal{F}$ to $\mathcal{F}^{c}$, for $i\in\mathbb{N}$.
The dwell time in the $i^{\text{th}}$ activation of the subsystems
$a$ and $u$ is defined as $\Delta t_{i}^{a}\triangleq t_{i}^{u}-t_{i}^{a}\in\mathbb{R}_{>0}$
and $\Delta t_{i}^{u}\triangleq t_{i+1}^{a}-t_{i}^{u}\in\mathbb{R}_{>0}$,
respectively. By Assumption \ref{assumption: x(0) in F} subsystem
$a$ is activated when $t=0$, and consequently $t_{i}^{u}>t_{i}^{a},\ \forall i\in\mathbb{N}$.
\begin{assumption}
The system is initialized in a feedback region (i.e. $x(0)\in\mathcal{F}$).
\label{assumption: x(0) in F}
\end{assumption}
To analyze the switched system, a common Lyapunov-like function is
designed as
\begin{equation}
V_{\sigma}(z(t))=V_{1}(e_{1}(t))+V_{2}(e_{2}(t)),\label{eq:switched_lyapunov}
\end{equation}
where the candidate Lyapunov functions for the tracking error and
the estimation error are selected respectively as

\begin{eqnarray}
V_{1}(e_{1}(t)) & = & \frac{1}{2}e_{1}^{T}(t)e_{1}(t),\label{eq:v1}\\
V_{2}(e_{2}(t)) & = & \frac{1}{2}e_{2}^{T}(t)e_{2}(t).\label{eq:v2}
\end{eqnarray}
The common Lyapunov-like function $V_{\sigma}(z(t))$ globally exponentially
converges while $x(t)\in\mathcal{F}$ and exhibits an exponential
growth when $x(t)\in\mathcal{F}^{c}$. Hence, a desired maximum bound
$V_{M}$ and a minimum threshold $V_{T}$ on $V_{\sigma}(z(t))$ may
be imposed such that $V_{\sigma}(z(t))\leq V_{M}$ and $V_{\sigma}(z(t_{i}^{u}))\leq V_{T}$.
A representative illustration for the evolution of $V_{\sigma}(z(t))$
is shown in Figure \ref{fig:evolution_of_v}. A lower threshold, $V_{T}$,
enforces the convergence of $\Vert z(t)\Vert$ to an arbitrary small
value. When implementing a high-frequency controller, $V_{T}$ may
be selected arbitrarily close to zero. However, the closer $V_{T}$
is selected to zero, the longer $x(t)$ is required to remain in $\mathcal{F}$,
and therefore the selection of $V_{T}$ is dependent of the individual
application tolerance. When a high-gain controller (e.g., $v_{r}(t)=k_{2}e_{2}(t)+\frac{\bar{d}^{2}}{\epsilon}e_{2}(t)$)
is implemented, $V_{T}$ should be selected such that $V_{T}\geq\epsilon$,
where $\epsilon$ is a design parameter.
\begin{thm}
\label{thm:The-switched-system}The composite error system trajectories
of the switched system generated by the family of subsystems described
by (\ref{eq:e1Dot_closed}), (\ref{eq:e2Dot closed loop}), and a
piecewise constant, right continuous switching signal $\sigma:[0,\infty)\rightarrow p\in\{a,u\}$
are globally uniformly ultimately bounded provided the switching signal
satisfies the minimum feedback availability dwell time condition

\begin{equation}
\Delta t_{i}^{a}\geq\frac{-1}{\lambda_{s}}\ln\left(\min\left(\frac{V_{T}}{V_{\sigma}(z(t_{i}^{a}))},\ 1\right)\right)\label{eq:minimum_observable_dwell_time_condition}
\end{equation}
and the maximum loss of feedback dwell time condition

\begin{equation}
\Delta t_{i}^{u}\leq\frac{1}{\lambda_{u}}\ln\left(\frac{V_{M}+\frac{\bar{d}^{2}}{2\lambda_{u}}}{V_{\sigma}(z(t_{i}^{u}))+\frac{\bar{d}^{2}}{2\lambda_{u}}}\right),\label{eq:maximum_unobservable_dwell_time_condition}
\end{equation}
where $\lambda_{s}$ and $\lambda_{u}$ are subsequently defined known
positive constants.
\end{thm}
\begin{IEEEproof}
By taking the time derivative of (\ref{eq:v1}) and substituting for
(\ref{eq:e1Dot_closed}) yields

\begin{equation}
\dot{V}_{1}(e_{1}(t))\leq-2\underbar{k}_{1}V_{1}(e_{1}(t)),\:\forall t,\label{eq:v1Dot}
\end{equation}
where $\underbar{k}_{1}$ is the minimum eigenvalue of $k_{1}$. By
using (\ref{eq:e2Dot closed loop}), the time derivative of (\ref{eq:v2})
can be expressed as

\begin{equation}
\dot{V}_{2}(e_{2}(t))\leq\begin{cases}
-2(\underbar{k}_{2}-c)V_{2}(e_{2}(t)), & t\in[t_{i}^{a},t_{i}^{u}),\\
\lambda_{u}V_{2}(e_{2}(t))+\frac{1}{2}\bar{d}^{2}, & t\in[t_{i}^{u},t_{i+1}^{a}),
\end{cases}\label{eq:v2Dot}
\end{equation}
where $c\in\mathbb{R}_{>0}$ is a Lipschitz constant, $\underbar{k}_{2}>c\in\mathbb{R}$
is the minimum eigenvalue of $k_{2}$, and $\lambda_{u}\triangleq2c+1\in\mathbb{R}_{>0}$.

From (\ref{eq:v1Dot}) and (\ref{eq:v2Dot}), the time derivative
of the common Lyapunov-like function can be expressed as
\begin{equation}
\dot{V}_{\sigma}(z(t))\leq\begin{cases}
-\lambda_{s}V_{\sigma}(z(t)), & t\in[t_{i}^{a},t_{i}^{u}),\\
\lambda_{u}V_{\sigma}(z(t))+\frac{1}{2}\bar{d}^{2}, & t\in[t_{i}^{u},t_{i+1}^{a}),
\end{cases}\;\forall i\in\mathbb{N},\label{eq:switched_lyapunov_derivative_cases}
\end{equation}
where $\lambda_{s}=2min\left(\underbar{k}_{1},\left(\underbar{k}_{2}-c\right)\right)\in\mathbb{R}_{>0}$.
The solutions to (\ref{eq:switched_lyapunov_derivative_cases}) for
the two subsystems are

\begin{align}
V_{\sigma}(z(t)) & \leq V_{\sigma}(z(t_{i}^{a}))e^{-\lambda_{s}\left(t-t_{i}^{a}\right)},\:t\in[t_{i}^{a},t_{i}^{u}),\label{eq:switched_lyapunov_solution_stable}\\
V_{\sigma}(z(t)) & \leq V_{\sigma}(z(t_{i}^{u}))e^{\lambda_{u}\left(t-t_{i}^{u}\right)}\nonumber \\
 & -\frac{\bar{d}^{2}}{2\lambda_{u}}\left(1-e^{\lambda_{u}\left(t-t_{i}^{u}\right)}\right),\:t\in[t_{i}^{u},t_{i+1}^{a}).\label{eq:switched_lyapunov_solution_unstable}
\end{align}

The inequality in (\ref{eq:switched_lyapunov_solution_stable}) indicates
that $\Vert z(t)\Vert\leq\Vert z(t_{i}^{a})\Vert e^{-\frac{1}{2}\lambda_{s}\left(t-t_{i}^{a}\right)},\ t\in[t_{i}^{a},t_{i}^{u})$
. The minimum threshold $V_{T}$ is selected to enforce the convergence
of $\Vert z(t)\Vert$ to desired threshold before allowing $x(t)$
to transition into $\mathcal{F}^{c}$. This condition can be expressed
as $V_{\sigma}(z(t_{i}^{a}))e^{-\lambda_{s}\Delta t_{i}^{a}}\leq V_{T}$,
and therefore the condition in (\ref{eq:minimum_observable_dwell_time_condition})
is obtained after algebraic manipulation. If $\frac{V_{T}}{V_{\sigma}(t_{i}^{a})}>1$,
the value of $V_{\sigma}(t_{i}^{a})$ is already below the threshold
and thus no minimum dwell time is required for the subsystem. 

When $t\in[t_{i}^{u},t_{i+1}^{a})$, the inequality in (\ref{eq:switched_lyapunov_solution_unstable})
indicates that $\Vert z\Vert\leq\sqrt{\Vert z(t_{i}^{u})\Vert^{2}e^{\lambda_{u}(t-t_{i}^{u})}-\frac{\bar{d}^{2}}{2\lambda_{u}}\left(1-e^{\lambda_{u}\left(t-t_{i}^{u}\right)}\right)}$,
and hence, the maximum bound $V_{M}$ is selected to limit the growth
of errors, where $V_{M}>V_{T}$. The maximum dwell time condition
for each of the $i^{\text{th}}$ unstable periods is expressed as
$V_{\sigma}(z(t_{i}^{u}))e^{\lambda_{u}\Delta t_{i}^{u}}-\frac{\bar{d}^{2}}{2\lambda_{u}}\left(1-e^{\lambda_{u}\Delta t_{i}^{u}}\right)\leq V_{M},$
and therefore the condition in (\ref{eq:maximum_unobservable_dwell_time_condition})
can be obtained.

Therefore, the composite error system trajectories generated by (\ref{eq:e1Dot_closed})
and (\ref{eq:e2Dot closed loop}) are globally uniformly ultimately
bounded as depicted in Figure \ref{fig:evolution_of_v}.
\end{IEEEproof}

\section{Switching Trajectory Design\label{sec:Trajectory-Design}}

Since $x_{d}$ lies outside the feedback region, i.e. $x_{d}\subset\mathcal{F}^{c},\ \forall t$,
and cannot be followed for all time, the switching trajectory $\bar{x}_{d}(t)$
is designed to enable $x(t)$ to follow $x_{d}$ to the extent possible
given the dwell time conditions in (\ref{eq:minimum_observable_dwell_time_condition})
and (\ref{eq:maximum_unobservable_dwell_time_condition}). A design
challenge for $\bar{x}_{d}(t)$ is to ensure $x(t)$ re-enters $\mathcal{F}$
to satisfy the sufficient condition in (\ref{eq:maximum_unobservable_dwell_time_condition}).
While $x(t)$ transitions through $\mathcal{F}^{c}$, $e(t)$ may
grow as indicated by (\ref{eq:switched_lyapunov_solution_unstable}),
and this growth must be accounted for when designing $\bar{x}_{d}(t)$.
To facilitate the development of the switching trajectory $\bar{x}_{d}(t)$,
$x_{b}(t)\in\mathbb{R}^{n}$ is defined as the closest orthogonal
projection of $\bar{x}_{d}(t)$ on the boundary of $\mathcal{F}$. 

When the maximum dwell time condition is reached, $\Vert e(t)\Vert\leq2\sqrt{V_{M}}$.
This bound implies there exist a set $\mathcal{B}=\left\{ y\in\mathbb{R}^{n}|\Vert y-\bar{x}_{d}(t)\Vert\leq2\sqrt{V_{M}}\right\} $
such that $x(t)\in\mathcal{B},\ \forall t$. Therefore, the switching
trajectory must penetrate a sufficient distance into $\mathcal{F}$
to compensate for the error accumulation. The distance to compensate
for error growth motivates the design of a cushion that ensures the
re-entry of the actual states when the maximum dwell time is reached.
Based on $x_{b}(t)$, the cushion $x_{\epsilon}(t)\in\mathbb{R}^{n}$
is selected as 

\begin{equation}
x_{\epsilon}(t)=x_{b}(t)+\Phi,\label{eq:secondary_trajectory}
\end{equation}
where $\Phi\in\mathbb{R}^{n}$, such that $\Vert\Phi\Vert\geq2\sqrt{V_{M}}$
and $\mathcal{B}\subseteq\mathcal{F}$. The general design rule is
that the switching trajectory must coincide with $x_{\epsilon}(t)$
when $t=t_{i}^{u}+\Delta t_{i}^{u}$, implying that $x(t)\in\mathcal{B}\subseteq\mathcal{F}$
when the maximum dwell time is reached.

\subsection{Design example}

An example switching trajectory $\bar{x}_{d}(t)$ can be developed
utilizing a smootherstep function described in \cite{Ebert2003} to
transition smoothly between $x_{d}$ and $x_{\epsilon}(t)$ while
meeting the dwell time conditions (see Remark \ref{rem:The-switching-trajectory-scheme}).
The smootherstep function is defined in \cite{Ebert2003} as 

\begin{eqnarray}
S(\rho) & = & 6\rho^{5}-15\rho^{4}+10\rho^{3}\label{eq:smootherstep}
\end{eqnarray}
where $\rho\in[0,1]$ is the input parameter. Given the transition
function in (\ref{eq:smootherstep}), the switching trajectory is
designed as 

\begin{align}
\bar{x}_{d}(t) & \triangleq\begin{cases}
H\Bigl(S(\rho_{i}^{a}),x_{b}(t),x_{\epsilon}(t)\Bigr), & t_{i}^{a}\leq t<t_{i}^{u},\\
H\Bigl(S(\rho_{i}^{u1}),g\left(x_{d},t\right),x_{b}(t)\Bigr), & t_{i}^{u}\leq t<t_{i}^{u1},\\
H\Bigl(S(\rho_{i}^{u2}),g\left(x_{d},t\right),g\left(x_{d},t\right)\Bigr), & t_{i}^{u1}\leq t<t_{i}^{u2},\\
H\Bigl(S(\rho_{i}^{u3}),x_{\epsilon}(t),g\left(x_{d},t\right)\Bigr), & t_{i}^{u2}\leq t<t_{i}^{u3},
\end{cases}\label{eq:switching_trajectory}
\end{align}
where $H\left(S(\cdot),q\left(t\right),r\left(t\right)\right)\triangleq S(\cdot)q\left(t\right)+\left[1-S(\cdot)\right]r\left(t\right)$
for $q(t),\ r(t)\in\mathbb{R}^{n},$ $g:x_{d}\times\mathbb{R}\rightarrow\mathbb{R}^{n}$
gives the desired state on $x_{d}$ at time $t$, $\rho_{i}^{a},\ \rho_{i}^{u1}\ \rho_{i}^{u2},\text{ and }\rho_{i}^{u3}$
are designed as $\rho_{i}^{a}\triangleq\frac{t-t_{i}^{a}}{\Delta t_{i}^{a}}$
and $\rho_{i}^{uj+1}\triangleq\frac{t-\left(t_{i}^{u}+\sum_{k=0}^{j}p_{k}\Delta t_{i}^{u}\right)}{p_{j+1}\Delta t_{i}^{u}},\ j\in\left\{ 0,1,2\right\} $,
the weights used to partition the maximum dwell time are denoted by
$p_{k}\in\left[0,1\right)$, and the corresponding partitions are
denoted by $t_{i}^{uj+1}$. The final partition, $t_{i}^{u3}$, coincides
with $t_{i+1}^{a}$. To avoid singularity in $\rho_{i}^{a}$ and to
ensure a smooth and continuous switching trajectory, $\Delta t_{i}^{a}$
must be arbitrarily lower bounded above zero (see Remark \ref{rem:Lower-bounding-ton}). 
\begin{rem}
Other trajectories satisfying the dwell time conditions in Theorem
\ref{thm:The-switched-system} may also be implemented, such as the
work in \cite{Chen.Bell.ea2017}.\label{rem:The-switching-trajectory-scheme}
\end{rem}
\begin{rem}
Lower bounding $\Delta t_{i}^{a}$ by an arbitrary value, $\alpha\in\mathbb{R}_{>0}$,
does not violate Theorem \ref{thm:The-switched-system} since the
system is allowed to remain in the feedback region longer than the
minimum dwell time, implying that $\Delta t_{i}^{a}\leq\alpha\leq\left(t-t_{i}^{a}\right)$
holds. Other trajectory designs may not require $\Delta t_{i}^{a}$
to be lower bounded.\label{rem:Lower-bounding-ton}
\end{rem}

\section{Simulation\label{sec:Simulation}}

A simulation is performed to illustrate the performance of the controller
given intermittent loss of state feedback. Based on the system model
given in (\ref{eq:qDot dynamics}), $f(x(t),t)$ is selected as $f(x(t),t)=Ax$
where $A=0.5I_{3}$, and $d(t)$ is drawn from a uniform distribution
between $\left[0,\ 0.06\right]$ meters per second. The initial states
and estimates are selected as $x(0)=\left[\begin{array}{ccc}
0.1\text{m} & 0.2\text{m} & 0\text{rads}\end{array}\right]$ and $\hat{x}(0)=\left[\begin{array}{ccc}
0.2\text{m} & 0.3\text{m} & \frac{\pi}{6}\text{rads}\end{array}\right]$. The observer and the controller gains were selected as$\begin{array}{cc}
k_{1}=3I_{3}\text{ and} & k_{2}=3I_{3},\end{array}$ respectively. The desired upper bound and lower threshold for the
composite error $\Vert z(t)\Vert$ are selected as 0.9 and 0.02 meters,
respectively. Based on the desired error bound and threshold, the
Lyapunov function bound and threshold are determined as $V_{M}=0.2025\text{ and }V_{T}=1\times10^{-4}$.

The desired path $x_{d}$ is selected as a circular trajectory with
a radius of 2 meters centered at the origin. The boundary of the feedback
region is selected as a circle with a 1-meter radius about the origin.
The switching trajectory $\bar{x}_{d}(t)$ were designed as described
in Section \ref{sec:Trajectory-Design} and follows $x_{d}$ at $\frac{\pi}{5}$
radians per second, where the partition weights are selected as$\begin{array}{cccc}
p_{0}0=0, & p_{1}=0.3, & p_{2}=0.4, & p_{3}=0.3\end{array}$.

\begin{figure}
\includegraphics[width=1\columnwidth]{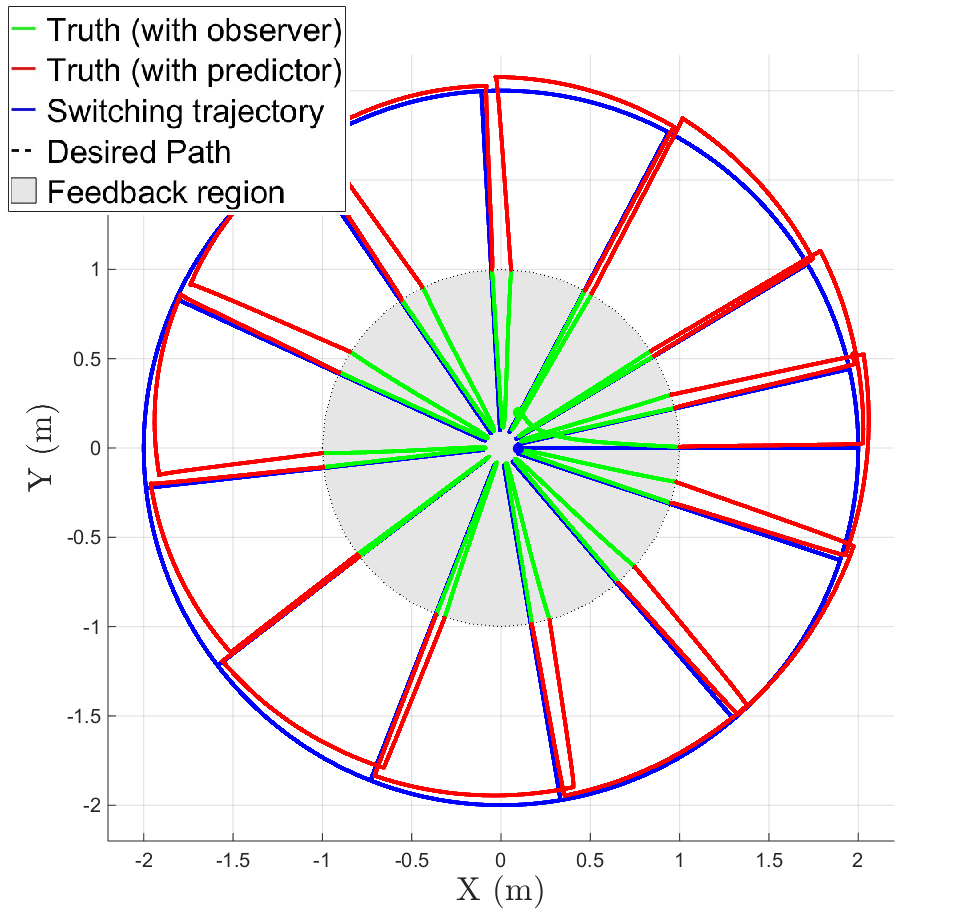}\caption{Simulation result for 30 seconds. Both system state $x(t)$ and switching
trajectory $\bar{x}_{d}(t)$ are initialized in the feedback region
(gray). During the minimum dwell time, $x(t)$ converges to $\bar{x}_{d}(t)$
exponentially with the observer activated. When $x(t)$ transitions
into the feedback-denied region (white), the predictor is activated,
and $x(t)$ gradually diverges from $\bar{x}_{d}(t)$ due to disturbances.
Before the maximum dwell time is reached, $x(t)$ re-enters the feedback
region and the observer is re-activated. Hence, $x(t)$ is able to
converge to $\bar{x}_{d}(t)$. \label{fig:Trajectory-tracking-result.-1}}
\end{figure}

\begin{figure}
\includegraphics[width=1\columnwidth]{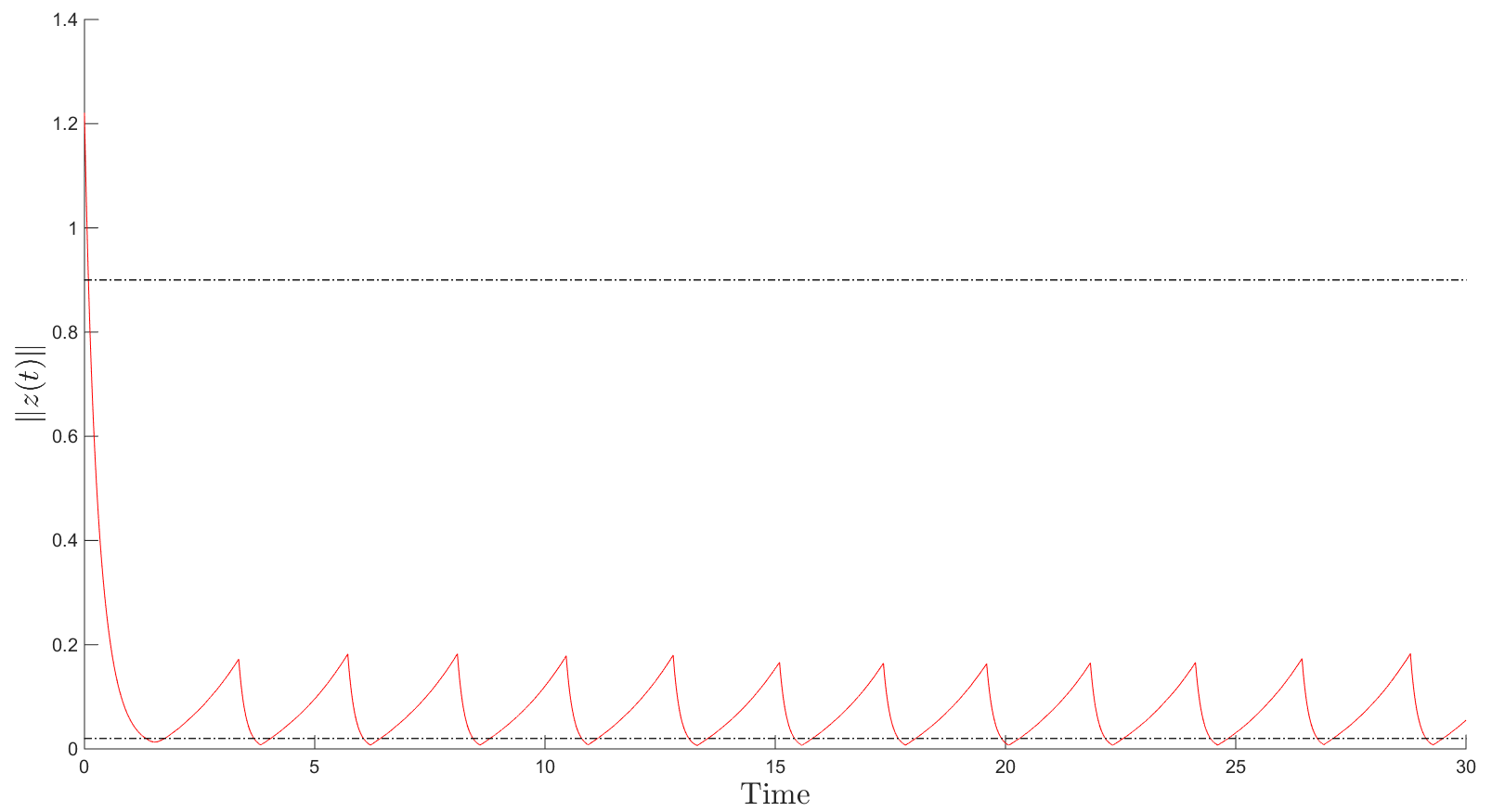}\caption{Evolution of $\Vert z(t)\Vert$. The top dashed line denotes $V_{M}$
and the bottom dashed line denotes $V_{T}$. \label{fig:State-estimate-tracking-1}}
\end{figure}
 Figure \ref{fig:Trajectory-tracking-result.-1} depicts the agent's
planar trajectory and shows that when the agent was inside the region
with state feedback, both the estimation and tracking errors, $\Vert e_{1}(t)\Vert$
and $\Vert e_{2}(t)\Vert$, exponentially converged. When the agent
was outside the feedback region, the tracking error converged while
the predictor error exhibited exponential divergence. 

The average maximum and minimum dwell times between switches are 2.16
and 0.26 seconds, respectively. Based on the simulation result, the
system is allowed to remain 8.23 times longer outside the feedback
region than inside on average. Furthermore, 40\% of the maximum dwell
time is dedicated to following the desired path, which translates
to 36\% of the combined duration of the maximum and minimum dwell
times per cycle. 

In Figure \ref{fig:State-estimate-tracking-1}, the composite error
$\Vert z(t)\Vert$ is shown. Figure \ref{fig:State-estimate-tracking-1}
indicates that $\Vert z(t)\Vert$ remained below 0.9 meters for all
time and less than or equal to 0.02 (indicated by the black dashed
line) by the end of each stable period, which demonstrates the robustness
of the presented control design under the dwell time condition constraints
and disturbances. Since an exact model of the system was used in this
simulation, the resulting tracking error is bounded well below the
maximum bound, and hence emphasizing the conservative nature of the
Lyapunov analysis method. 

\section{Experiments\label{sec:Experiments}}

In Section \ref{sec:Experiments}, an experiment is performed to verify
the theoretical results where a single integrator dynamic is used
instead of the exact system model. The overall goal of the experiment
is to represent a scenario where an unmanned air vehicle is tasked
with following a path where feedback is not available (e.g., inside
an urban canyon). Specifically, the objective is to demonstrate the
boundedness of the tracking error $e(t)$ through multiple cycles
of switching between the feedback-available and unavailable regions
based on the dwell time constraints established in Section \ref{sec:Switched-System-Analysis}.
A Parrot Bebop 2.0 quadcopter is used as the unmanned air vehicle.
The quadcopter is equipped with a 3-axis gyroscope, a 3-axis accelerometer,
an ultrasound sensor, and an optical-flow sensor. The on-board sensors
provide an estimate of the linear and angular velocities of the quadcopter
at 5Hz. To control the quadcopter, the \textit{bebop\_autonomy} package
developed by \cite{BEBOP_AUTONOMYlibrary} is utilized to send velocity
commands generated from an off-board computer running Robotic Operating
System (ROS) Kinetic in Ubuntu 16.04. The communication link between
the computer and the quadcopter is established through a WiFi channel
at 5GHz. 

A NaturalPoint, Inc. OptiTrack motion capture system is used to simulate
a feedback signal and record the ground truth pose of the quadcopter
at a rate of 120Hz. While the quadcopter is inside the feedback region,
pose information from the motion capture system is directly used as
feedback in the controller and update laws designed in Section \ref{sec:Controller}.
When the quadcopter operated outside of the feedback region, the pose
feedback is discarded. During these times, the on-board velocity measurements
are used to feedforward the state estimate. Although the OptiTrack
system continue to record the pose of the quadcopter, the pose information
is only used as ground truth for illustration purposes.

Utilizing the motion capture system, a circular region of available
feedback is centered at the origin of the Euclidean world frame with
a radius of 1 meter. Since torque level control authority is not available,
single integrator dynamics, $\dot{q}(t)=u(t)+d(t)$, are assumed for
the quadcopter where $q(t)=\left[\begin{array}{cccc}
x(t) & y(t) & z(t) & \alpha(t)\end{array}\right]^{T}$, and $x(t),\ y(t),\ z(t),\ \alpha(t)\in\mathbb{R}$ are the 3-D Euclidean
coordinates and yaw rotation of the quadcopter with respect to the
inertial frame. The disturbance is assumed to be upper bounded as
$\bar{d}=0.035$. To compensate for the disturbance, a high-gain robust
controller is implemented to ensure a continuous control command.
The controller and update law gains are selected as $k_{1}=0.4I_{3},\ k_{2}=0.6I_{3},\text{ and }\epsilon=0.1$.
To regulate and match the actual velocity output to the control command,
a low level PID controller is implemented. 

The desired upper bound and lower threshold on $\Vert z(t)\Vert$
are selected as 0.9 and 0.14 meters, respectively. Since single integrator
dynamics are assumed for the quadcopter dynamic, a less conservative
minimum dwell time condition can be derived (details are given in
the Appendix). The desired path is defined as a circular path centered
at the origin with a radius of 1.5 meters. Following the design method
outlined in Section \ref{sec:Trajectory-Design}, a switching trajectory
is designed to follow $x_{d}$ with an angular velocity of $\frac{\pi}{15}$
radians per second. To prevent the quadcopter from drifting out of
the feedback region prematurely, a intermediate trajectory is design
to be $x_{int}(t)=0.7x_{b}(t)$ to replace $x_{b}(t)$ in (\ref{eq:switching_trajectory})
as a safety measure. The partitions for the maximum dwell time are
selected as $p_{0}=0,\ p_{1}=0.4,\ p_{2}=0.2,\ p_{3}=0.4$.

\begin{figure}
\includegraphics[width=1\columnwidth]{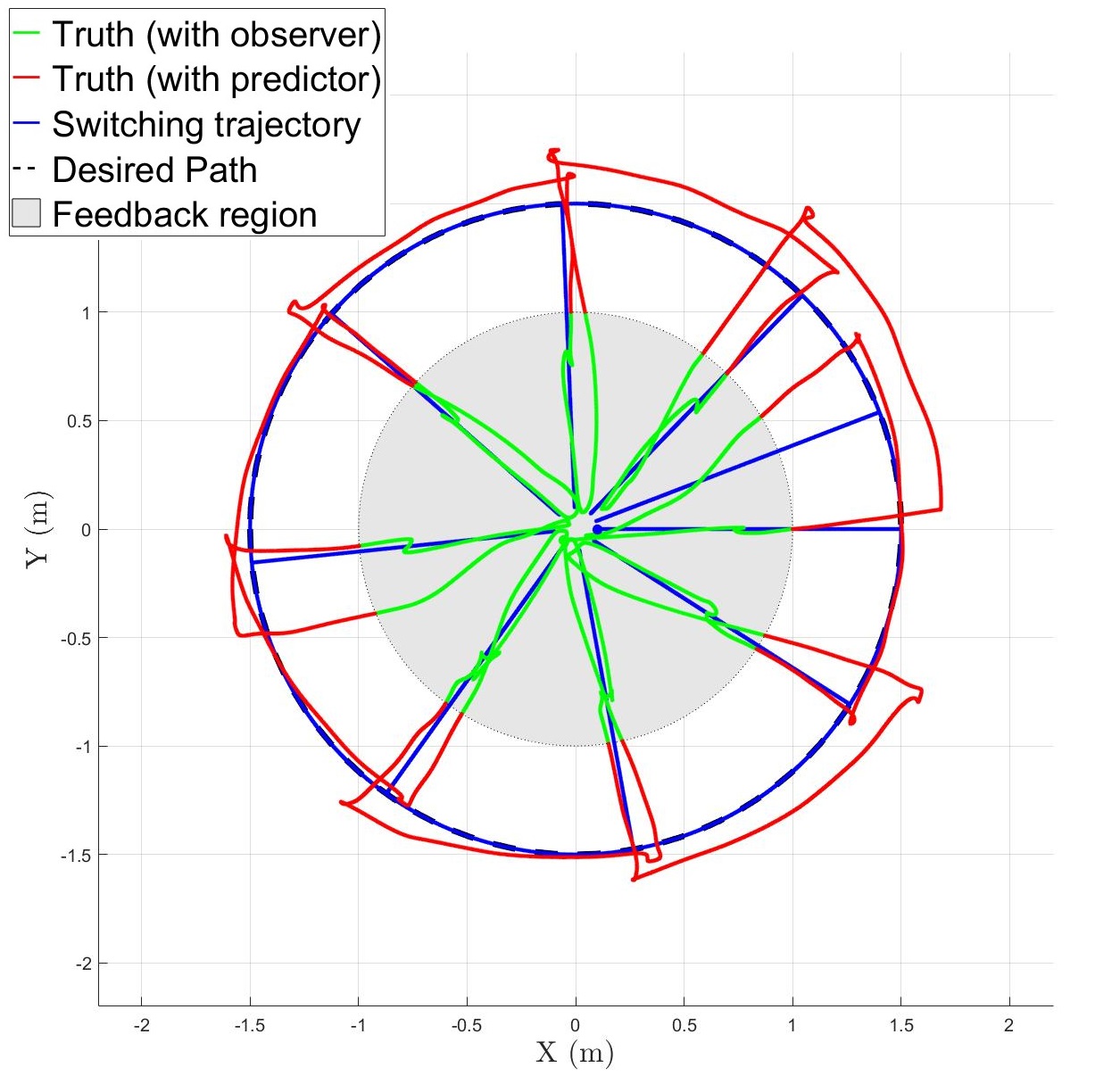}\caption{Actual and switching trajectory over 185 seconds.\label{fig:tracking_exp}}
\end{figure}

Initially, the quadcopter is launched inside $\mathcal{F}$ along
with the switching trajectory, which transitions between $\mathcal{F}$
and $x_{d}$ over the prescribed time span. The experimental results
demonstrate that the quadcopter is capable of intermittently leaving
$\mathcal{F}$ to follow $x_{d}$ for some period of time and then
return to $\mathcal{F}$ consistently. The supplementary video accompanying
this paper, available for download at http://ieeexplore.ieee.org,
gives a recording of the experiment with the motion of the quadcopter
and the switching trajectory projected on the floor. The overall path
following plot, including the desired path, switching trajectory and
actual states, is shown in Figure \ref{fig:tracking_exp}, where a
total of 8 cycles of leaving and re-entering $\mathcal{F}$ occurred.
During the periods when the quadcopter is outside the feedback region,
large odometry drifts are apparent and the actual tracking error diverges
as the dynamic models in Section \ref{sec:Switched-System-Analysis}
indicate. Table \ref{tab:DT-table-exp} indicates the maximum and
minimum dwell times for each cycle. On average, the quadcopter was
allowed to reside approximately 6 times longer in $\mathcal{F}^{c}$
than $\mathcal{F}$, and 20\% of which is dedicated to following $x_{d}$.
Specifically, the quadcopter is allowed 19.85 seconds in $\mathcal{F}^{c}$
and is required to remain in $\mathcal{F}$ for 3.31 seconds on average.
Based on the partition weights of the maximum dwell time, Table \ref{tab:max_DT-partitions-exp}
describes the partitions and the duration for each partition. During
partition 1, $\bar{x}_{d}(t)$ transitions from the $x_{b}(t)$ to
$x_{d}$ where the partition weight was set to 50\%. The relatively
large partition allots more time in transition to yield a slower velocity
profile, which produces less overshoot in the tracking performance.
The distance between $x_{d}$ and $\mathcal{F}$ is also a major factor
in distributing partition weights in the sense that the closer $x_{d}$
is to $\mathcal{F}$, the less time is required for transition and
more time can be allocated to follow $x_{d}$. 

\begin{table}
\centering{}\caption{Minimum and Maximum Dwell Times.\label{tab:DT-table-exp}}
\begin{tabular}{|c|c|c|}
\hline 
Cycle &
Max. D. T. (s) &
Min. D. T. (s)\tabularnewline
\hline 
\hline 
0 &
- &
3.50\tabularnewline
\hline 
1 &
19.12 &
4.55\tabularnewline
\hline 
2 &
19.38 &
4.09\tabularnewline
\hline 
3 &
19.25 &
3.20\tabularnewline
\hline 
4 &
19.72 &
3.34\tabularnewline
\hline 
5 &
20.21 &
1.67\tabularnewline
\hline 
6 &
19.08 &
2.55\tabularnewline
\hline 
7 &
19.65 &
3.73\tabularnewline
\hline 
8 &
22.35 &
3.16\tabularnewline
\hline 
Avg &
19.85 &
3.31\tabularnewline
\hline 
\end{tabular}
\end{table}

\begin{table}
\noindent \centering{}\caption{Maximum Dwell Time Partitions.\label{tab:max_DT-partitions-exp}}
\begin{tabular}{|c|c|c|c|}
\hline 
\multirow{2}{*}{Cycle} &
\multicolumn{3}{c|}{Maximum dwell times (s)}\tabularnewline
\cline{2-4} 
 & Part. 1 (40\%) &
Part. 2 (20\%) &
Part. 3 (40\%)\tabularnewline
\hline 
\hline 
1 &
7.65 &
3.82 &
7.65\tabularnewline
\hline 
2 &
7.75 &
3.88 &
7.75\tabularnewline
\hline 
3 &
7.70 &
3.85 &
7.70\tabularnewline
\hline 
4 &
7.89 &
3.94 &
7.89\tabularnewline
\hline 
5 &
8.08 &
4.04 &
8.08\tabularnewline
\hline 
6 &
7.63 &
3.82 &
7.63\tabularnewline
\hline 
7 &
7.86 &
3.93 &
7.86\tabularnewline
\hline 
8 &
8.94 &
4.47 &
8.94\tabularnewline
\hline 
\end{tabular}
\end{table}

\begin{figure}

\includegraphics[width=1\columnwidth]{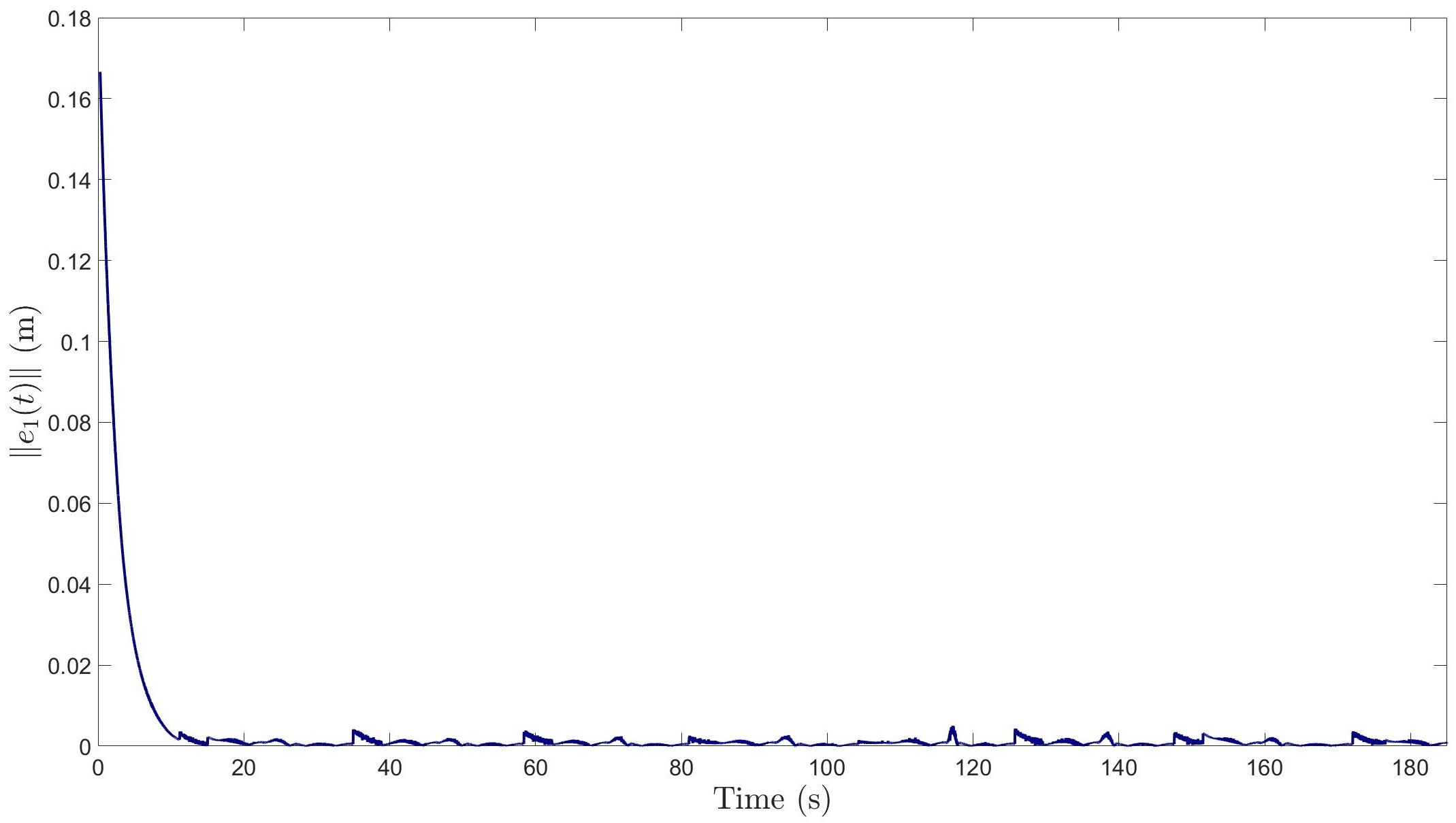}

\caption{Estimate tracking error $\Vert e_{1}(t)\Vert$. As indicated by the
analysis, the estimate tracking error exhibits exponential stability
regardless of feedback availability.\label{fig:Estimate-tracking-error-exp}}
\end{figure}

\begin{figure}
\includegraphics[width=1\columnwidth]{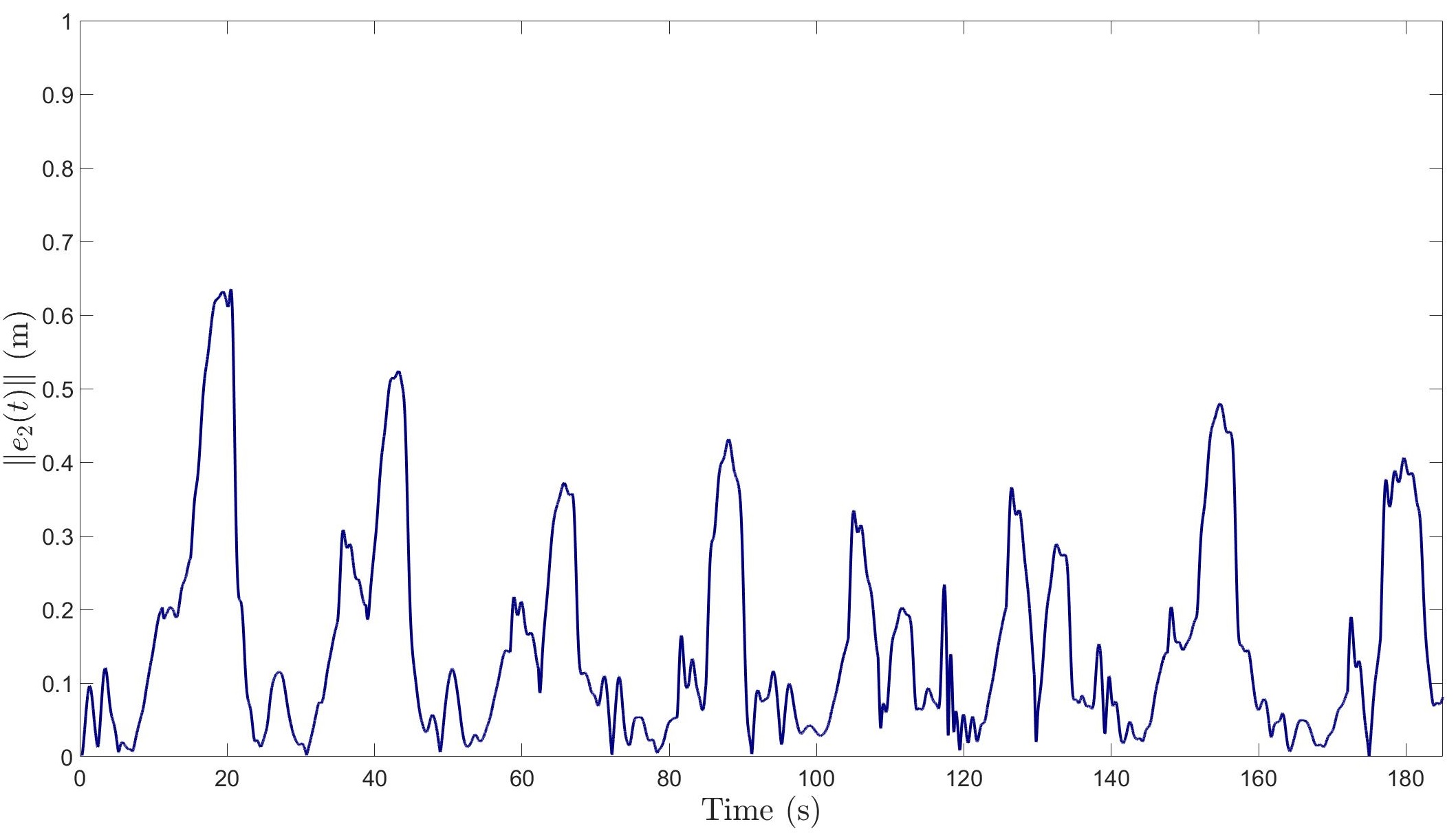}\caption{Estimation error $\Vert e_{2}(t)\Vert$. As indicated by the analysis,
the estimation error converges when $x(t)\in\mathcal{F}$ and diverges
when $x(t)\in\mathcal{F}^{c}$. \label{fig:Estimation-error-exp}}
\end{figure}

\begin{figure}
\includegraphics[width=1\columnwidth]{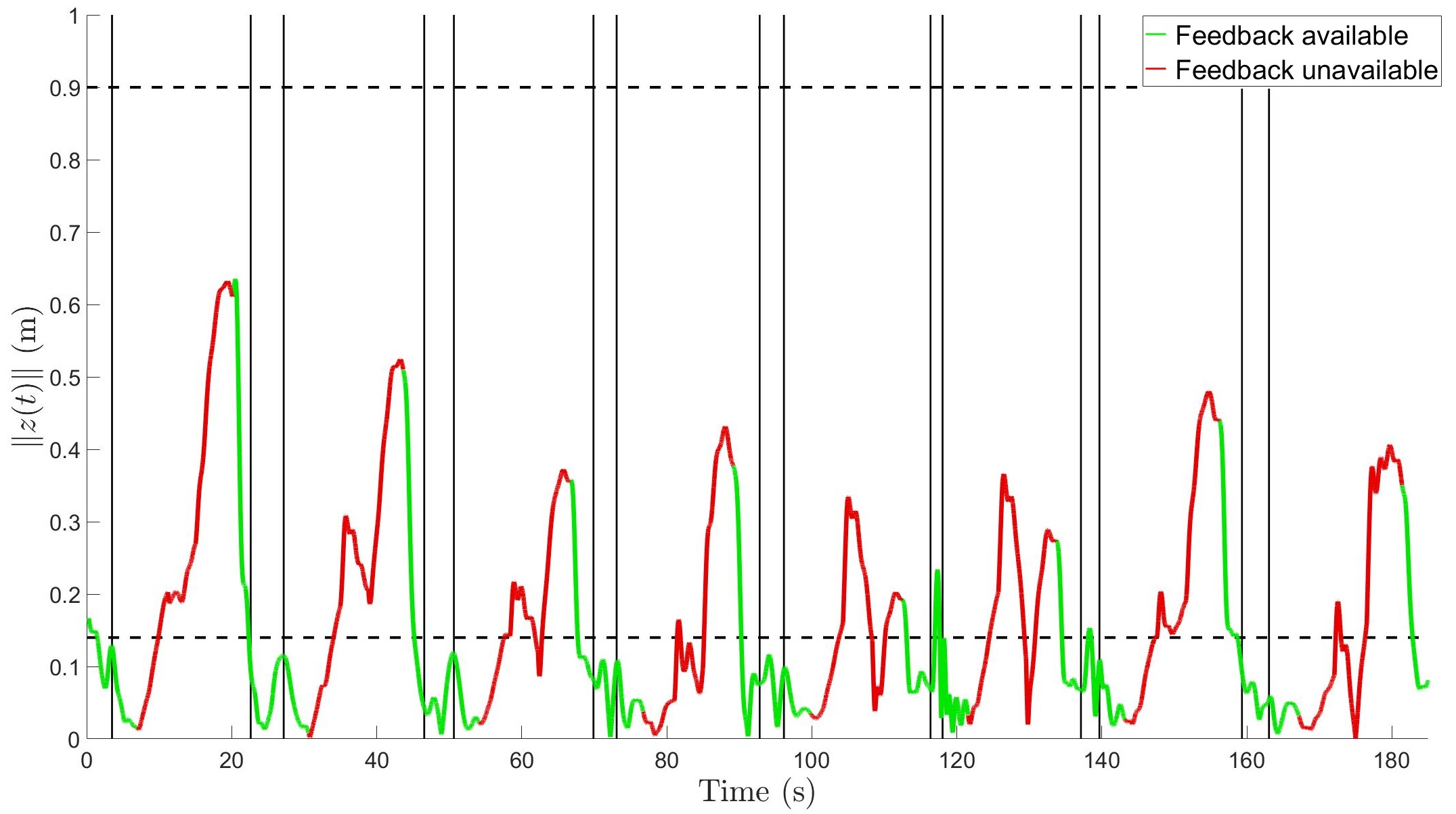}\caption{Evolution of $\Vert z(t)\Vert$. The dash-dot (vertical) lines indicate
the switching interface of minimum and maximum dwell times, and the
dashed (horizontal) lines indicate the prescribed upper bound and
lower threshold.\label{fig:z_norm-exp}}
\end{figure}

\begin{figure}
\includegraphics[width=1\columnwidth]{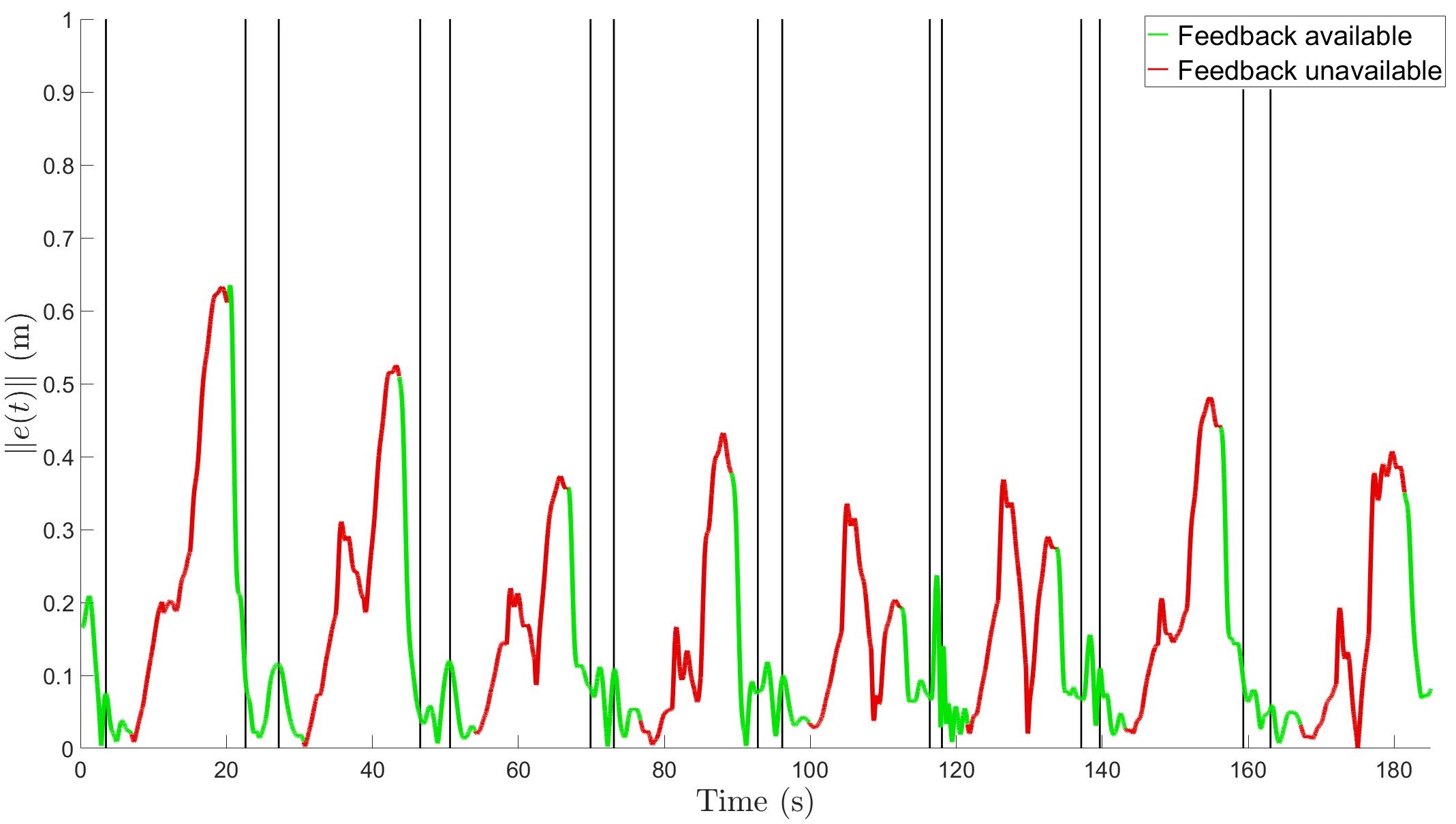}\caption{Actual tracking error $\Vert e(t)\Vert$. The dash-dot lines indicate
the switching interface of minimum and maximum dwell times. \label{fig:Actual-tracking-error-exp}}
\end{figure}

\begin{figure}
\includegraphics[width=1\columnwidth]{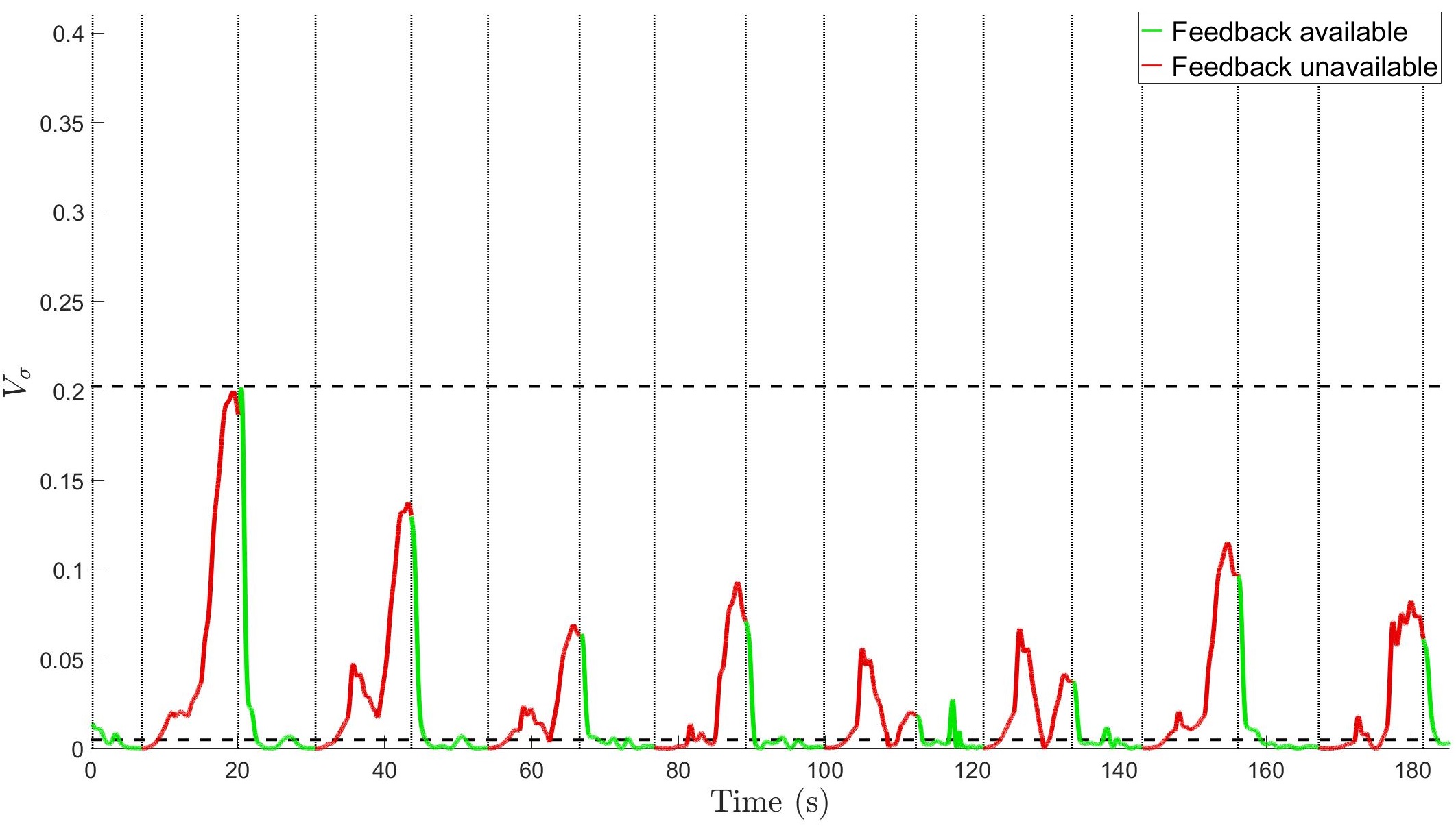}\caption{Evolution of $V_{\sigma}(t)$. The dotted (vertical) lines indicate
the time instants when the quadcopter crossed the feedback region
boundary. The dashed (horizontal) lines indicate the prescribed $V_{M}$
and $V_{T}$ for $V_{\sigma}$.\label{fig:v_sigma-exp}}

\end{figure}

To illustrate the stability of the control scheme, the Euclidean norm
of the estimate tracking error, $e_{1}(t)$, and the estimation error,
$e_{2}(t)$, are displayed in Figure \ref{fig:Estimate-tracking-error-exp}
and \ref{fig:Estimation-error-exp}. The estimate tracking error exponentially
converges, reflecting the analysis in (\ref{eq:v1Dot}). The estimation
error exhibits growth when $x(t)\in\mathcal{F}^{c}$. For a better
illustration, the norm of the composite and actual tracking error
are shown in Figure \ref{fig:z_norm-exp} and \ref{fig:Actual-tracking-error-exp},
respectively, where the dwell time duration is indicated by vertical
dash-dot lines and the upper bound and lower threshold on the actual
tracking error are indicated by horizontal dashed lines. Over the
8 cycles, $\Vert z(t)\Vert$ is upper bounded by 0.9 meters at all
times, and converges to below 0.14 meters within the minimum dwell
time when $x(t)\in\mathcal{F}$. The plots also indicate that $x(t)$
is able to return to $\mathcal{F}$ within the maximum dwell times.
This can be verified by the activation of the observer before the
maximum dwell time is reached for every cycle. In Figure \ref{fig:v_sigma-exp},
the evolution of $V_{\sigma}$ is shown along with the calculated
$V_{M}$ and $V_{T}$ as indicated by the horizontal dashed lines.
As expected, the Lyapunov-like function $V_{\sigma}$ is upper bounded
below $V_{M}$ for all times and converges below $V_{T}$ within the
minimum dwell times. Based on Figure \ref{fig:Actual-tracking-error-exp}
and \ref{fig:v_sigma-exp}, the controller and update laws developed
in Section \ref{sec:Controller} demonstrate robustness towards disturbances
and a simple assumed dynamic model. Hence, the trajectory design scheme
provided in Section \ref{sec:Trajectory-Design} is able to generate
a switching signal $\sigma(t)$ that satisfied the dwell time conditions
developed in Section \ref{sec:Switched-System-Analysis} and, therefore,
verifying the claim in Theorem \ref{thm:The-switched-system}.



\section{Conclusion\label{sec:Conclusion}}

A novel method that utilizes a switched systems approach to ensure
path following stability under intermittent state feedback is presented.
The presented method relieves the requirement of state feedback at
all times. State estimates are used in the tracking control to compensate
for the intermittence of state feedback. A Lyapunov-based, switched
systems analysis is used to develop maximum and minimum dwell time
conditions to guarantee stability of the overall system. The dwell
time conditions allow the desired path to be completely outside of
the feedback region, and a switching trajectory is designed to bring
the states back into the feedback region before the error growth exceeds
a defined threshold. The candidate switching trajectory switches between
the desired path and the feedback region using smootherstep transition
functions. A simulation and an experiment were performed to illustrate
the robustness of the control and trajectory design. Future research
will focus on development of an approximate optimal control approach
using adaptive dynamic programming concepts to yield approximately
optimal results. Further efforts will also examine cases where the
feedback region is time-varying or unknown.

\appendix{}

When using single integrator dynamics, $\dot{x}(t)=u+d(t)$, the resulting
estimation error dynamics for the unstable subsystem is $\Vert\dot{e}_{2}(t)\Vert\leq\bar{d}$
, and the corresponding Lyapunov-like function derivative is $\dot{V}_{\sigma}(t)\leq\bar{d}\Vert e_{2}(t)\Vert$.
By solving the ordinary differential equation for $\dot{e}_{2}(t)$,
the estimation error $e_{2}(t)$ exhibits a linear growth that can
be bounded as $e_{2}(t)\leq e_{2}(t_{i}^{u})+\bar{d}(t-t_{i}^{u})$.
After substituting in the linear bound on $e_{2}(t)$, it follows
that $\dot{V}_{\sigma}(t)\leq\bar{d}\Vert e_{2}(t_{i}^{u})\Vert+\bar{d}^{2}(t-t_{i}^{u})$,
and solving the ordinary differential equation yields $V_{\sigma}(t)\leq\frac{1}{2}\bar{d}^{2}\left(t-t_{i}^{u}\right){}^{2}+\bar{d}\Vert e_{2}(t_{i}^{u})\Vert\left(t-t_{i}^{u}\right)+V_{\sigma}(z(t_{i}^{u}))$.
After imposing $V_{\sigma}(t)\leq V_{M}$ as the upper bound constraint,
the maximum dwell time can be derived by solving the quadratic equation
and taking the positive root as 

\[
\Delta t_{i}^{u}\leq\frac{\left(\sqrt{\Vert e_{2}(t_{i}^{u})\Vert^{2}-2\left(V_{\sigma}(z(t_{i}^{u}))-V_{M}\right)}-\Vert e_{2}(t_{i}^{u})\Vert\right)}{\bar{d}}.
\]

\bibliographystyle{IEEEtran}
\bibliography{9C__NCR_ncr_bibtex_bib_ncrbibs_encr,10C__NCR_ncr_bibtex_bib_ncrbibs_master,11C__NCR_ncr_bibtex_bib_ncrbibs_ncr,IEEEabrv}

\end{document}